\documentclass[conference]{IEEEtran}

\usepackage[pdftex]{hyperref}
\usepackage{graphicx}
\usepackage{amsmath}
\usepackage{amssymb}
\usepackage{color}
\usepackage{ifpdf}

\usepackage{float}
\usepackage[utf8]{inputenc}
\usepackage{multirow}
\usepackage{rotating}
\usepackage{subfigure}

\usepackage{moresize}
\usepackage{url}
\usepackage{booktabs}
\usepackage{listings}   
\usepackage{paralist}    
\usepackage{wrapfig}    
\usepackage{multirow}
\usepackage{ifpdf}
\usepackage{xspace}
\usepackage{keyval}  
\usepackage{color}

\definecolor{listinggray}{gray}{0.95}
\definecolor{darkgray}{gray}{0.7}
\definecolor{commentgreen}{rgb}{0, 0.4, 0}
\definecolor{darkblue}{rgb}{0, 0, 0.4}
\definecolor{middleblue}{rgb}{0, 0, 0.7}
\definecolor{darkred}{rgb}{0.4, 0, 0}
\definecolor{brown}{rgb}{0.5, 0.5, 0}

\usepackage[normalem]{ulem}
\makeatletter
\def\cyanuwave{\bgroup \markoverwith{\lower3.5\p@\hbox{\sixly \textcolor{cyan}{\char58}}}\ULon}
\def\reduwave{\bgroup \markoverwith{\lower3.5\p@\hbox{\sixly \textcolor{red}{\char58}}}\ULon}
\def\blueuwave{\bgroup \markoverwith{\lower3.5\p@\hbox{\sixly \textcolor{blue}{\char58}}}\ULon}
\font\sixly=lasy6 
\makeatother

\usepackage{soul}
\usepackage{ulem}

\newif\ifdraft
\ifdraft
\usepackage{xcolor}
\definecolor{ocolor}{rgb}{1,0,0.4}
\newcommand{\onote}[1]{ {\textcolor{ocolor} { (***Ole: #1) }}}
\newcommand{\terminology}[1]{ {\textcolor{red} {(Terminology used: \textbf{#1}) }}}

\newcommand{\jhanote}[1]{ {\textcolor{red} { ***shantenu: #1 }}}
\newcommand{\alnote}[1]{ {\textcolor{blue} { ***andreL: #1 }}}
\newcommand{\amnote}[1]{ {\textcolor{blue} { ***andreM: #1 }}}
\newcommand{\smnote}[1]{ {\textcolor{brown} { ***sharath: #1 }}}
\newcommand{\pmnote}[1]{ {\textcolor{brown} { ***Pradeep: #1 }}}
\newcommand{\msnote}[1]{ {\textcolor{cyan} { ***mark: #1 }}}
\newcommand{\mrnote}[1]{ {\textcolor{purple} { ***melissa: #1 }}}
\definecolor{orange}{rgb}{1,.5,0}
\newcommand{\aznote}[1]{ {\textcolor{orange} { ***ashley: #1 }}}
\definecolor{dandelion}{cmyk}{0,0.29,0.84,0}
\newcommand{\mtnote}[1]{ {\textcolor{dandelion} { ***matteo: #1 }}}
\newcommand{\note}[1]{ {\textcolor{magenta} { ***Note: #1 }}}
\else
\newcommand{\onote}[1]{}
\newcommand{\terminology}[1]{}

\newcommand{\alnote}[1]{}
\newcommand{\amnote}[1]{}
\newcommand{\athotanote}[1]{}
\newcommand{\smnote}[1]{}
\newcommand{\pmnote}[1]{}
\newcommand{\jhanote}[1]{}
\newcommand{\msnote}[1]{}
\newcommand{\mrnote}[1]{}
\newcommand{\aznote}[1]{}
\newcommand{\mtnote}[1]{}
\newcommand{\note}[1]{}
\fi

\newcommand{\pilot}{Pilot\xspace}
\newcommand{\pilots}{Pilots\xspace}
\newcommand{\pilotjob}{Pilot-Job\xspace}
\newcommand{\pilotjobs}{Pilot-Jobs\xspace}

\newcommand{\pilotdata}{Pilot-Data\xspace}

\newcommand{\cu}{CU\xspace}

\newcommand{\upp}{\vspace*{-0.5em}}

\newcommand{\irods}{iRODS\xspace}

\lstdefinestyle{myListing}{
  frame=single,   
  backgroundcolor=\color{listinggray},  
  language=C,       
  basicstyle=\ttfamily \footnotesize,
  breakautoindent=true,
  breaklines=true
  tabsize=2,
  captionpos=b,  
  aboveskip=0em,
  belowskip=-2em,
}      

\lstdefinestyle{myPythonListing}{
  frame=single,   
  backgroundcolor=\color{listinggray},  
  language=Python,       
  basicstyle=\ttfamily \scriptsize,
  breakautoindent=true,
  breaklines=true
  tabsize=2,
  captionpos=b,  
}

\ifpdf
\DeclareGraphicsExtensions{.pdf, .jpg, .tif}
\else
\DeclareGraphicsExtensions{.ps,  .eps, .jpg}
\fi

\usepackage{listings}
\usepackage{paralist}
\lstnewenvironment{code}[1][]%
{
\noindent
\minipage{1.0 \linewidth} op
\vspace{0.5\baselineskip}
\lstset{
    language=Python,
    frame=single,
    captionpos=b,
    stringstyle=\ttfamily,
    basicstyle=\scriptsize\ttfamily,
    showstringspaces=false,#1}
}
{\endminipage}

\defaultleftmargin{1em}{}{}{}
\begin{document}

\title{A Tale of Two Data-Intensive Paradigms: Applications,
  Abstractions, and Architectures}

\author{Shantenu Jha$^{1}$, Judy Qiu$^{2}$, Andre Luckow$^{1}$,
  Pradeep Mantha$^{1}$, Geoffrey C.Fox$^{2}$\\
{\emph{$^{(1)}$ \footnotesize{RADICAL, Rutgers University, Piscataway,
      NJ 08854, USA}} \footnotesize{$^{(2)}$ Indiana University, USA}}\\
}

\date{}
\maketitle

\begin{abstract} 
  Scientific problems that depend on processing large amounts of data
  require overcoming challenges in multiple areas: managing
  large-scale data distribution, co-placement and scheduling of data
  with compute resources, and storing and transferring large volumes
  of data. We analyze the ecosystems of the two prominent paradigms
  for data-intensive applications, hereafter referred to as the
  high-performance computing and the Apache-Hadoop paradigm. We
  propose a basis, common terminology and functional factors upon
  which to analyze the two approaches of both paradigms. We discuss
  the concept of ``Big Data Ogres'' and their facets as means of
  understanding and characterizing the most common application
  workloads found across the two paradigms. We then discuss the
  salient features of the two paradigms, and compare and contrast the
  two approaches.  Specifically, we examine common
  implementation/approaches of these paradigms, shed light upon the
  reasons for their current ``architecture'' and discuss some typical
  workloads that utilize them. In spite of the significant software
  distinctions, we believe there is architectural similarity. We
  discuss the potential integration of different implementations,
  across the different levels and components. Our comparison
  progresses from a fully qualitative examination of the two
  paradigms, to a semi-quantitative methodology.  We use a simple and
  broadly used Ogre (K-means clustering), characterize its performance
  on a range of representative platforms, covering several
  implementations from both paradigms. Our experiments provide an
  insight into the relative strengths of the two paradigms.  We
  propose that the set of Ogres will serve as a benchmark to evaluate
  the two paradigms along different dimensions.


\end{abstract}

\section{Introduction} 

\note{Approach: Technology and Architecture, Paradigm: Approach + social aspects; Hadoop ecosystem for calling out specific Hadoop elements}

The growing importance of data-intensive applications is generally recognized
and has lead to a wide range of approaches and solutions for data distribution,
management and processing. These approaches are characterized by a broad set of
tools, software frameworks and implementations. Although seemingly unrelated,
the approaches can be better understood by examining their use of common
abstractions and similar architectures for data management and processing.
Building upon this putative similarly, we examine and organize many existing
approaches to Big Data processing into two primary paradigms -- the scientific
high-performance (HPC) and Apache Big Data Stack (ABDS) centered around Apache
Hadoop, which we believe reflects and captures the dominant historical,
technical and social forces that have shaped the landscape of Big Data
analytics.

The HPC paradigm has its roots in supercomputing-class computationally
intensive scientific problems (e.g. Molecular Dynamics of
macromolecular systems, fluid dynamics at scales to capture
turbulence) and in managing large-scale distributed problems (e.\,g.\
data analysis from the LHC). HPC paradigm has been characterized by
limited implementations, but customized and tuned for performance
along a narrow set of requirements.  In contrast, the Apache Big Data Stack
(ABDS) has seen a significant update in industry and recently also in
scientific environments. A vibrant, manifold open-source ecosystem
consisting of higher-level data stores, data processing/analytics and
machine learning frameworks has evolved around a stable,
non-monolithic kernel: the Hadoop Filesystem (HDFS) and YARN.  Hadoop
integrates compute and data, and introduces application-level
scheduling to facilitate heterogeneous application
workloads and high cluster utilization.

\note{Call out up front: 2 main differences between Hadoop and HPC: tight compute/data integration, multi-level scheduling}

The success and evolution of ABDS into a widely deployed
cluster computing frameworks yields many opportunities for {\it
  traditional} scientific applications; it also raises many important
questions: What features of ABDS are useful for traditional
scientific workloads?  What features of the ABDS can be
extended and integrated with HPC?  How do typical
data-intensive HPC and ABDS workloads differ?  It is
currently difficult for most applications to utilize the two paradigms
interoperably.  The divergence and heterogeneity will likely increase
due to the continuing evolution of ABDS, thus we believe it
is important and timely to answer questions that will support
interoperable approaches. However, before such interoperable
approaches can be formulated, it is important to understand the
different abstractions, architectures and applications that each
paradigm utilizes and supports.  The aim of this paper is to
provide the conceptual framework and terminology and to begin
addressing questions of interoperability.

{\it Paper Outline: } This paper is divided into two logical parts: in
the first, we analyze the ecosystem of the two primary paradigms to
data-intensive applications.
We discuss the salient features of the two paradigms,
compare and contrast the two for functionality and implementations
along the layers of analysis, runtime-environments, communication
layer, resource management layer and the physical resource layer. In
the second part, we move from a fully qualitative examination of the
two, to a semi-quantitative methodology, whereby we experimentally
examine both hard performance numbers (along different implementations
of the two stacks) and soft issues such as completeness, expressivity,
extensibility as well as software engineering
considerations.  

\upp
\section{Data-Intensive Application: Big Data Ogres} \label{sec:application_characteristics} 
\upp 
Based upon an analysis of a large set of Big Data applications, including more
than 50 use cases~\cite{nist-uc}, we propose the Big Data Ogres in analogy with
parallel computing with the Berkeley Dwarfs, NAS benchmarks and linear algebra
templates. The purpose of Big Data Ogres is to discern commonalities and
patterns across a broad range of seemingly different Big Data applications,
propose an initial structure to classify them, and help cluster some commonly
found applications using structure. Similar to the Berkeley Dwarfs, the Big
Data Ogres are not orthogonal, nor exclusive, and thus do not constitute a
formal taxonomy. We propose the Ogres as a benchmark to investigate and
evaluate the paradigms for architectural principles, capabilities and
implementation performance. Also, we capture the richness of Big Data by
including not just different parallel structures but also important overall
patterns. Big Data is in its infancy without clear consensus as to important
issues and so we propose an inclusive set of Ogres expecting that further
discussion will refine them.

\alnote{Are we talking about characteristics of data stores:
  file-based vs. relational vs. nosql (column, memory) etc.?}

The first facet captures {\bf different problem architectures}. Some
representative examples are (i) pleasingly parallel -- as in Blast
(over sequences), protein docking (over proteins and docking sites),
imagery, (ii) local machine learning (ML) -- or filtering pleasingly
parallel as in bio-imagery, radar (this contrasts global machine
learning seen in LDA, clustering etc. with parallel ML over nodes of
system), (iii) fusion -- where knowledge discovery often
involves fusion of multiple methods (ensemble methods are one
approach), (iv) data points in metric or non-metric spaces, (v)
maximum likelihood, (vi) ${\chi}^2$minimizations, (vii) expectation
maximization (often steepest descent), and (viii) quantitative
measures for Big Data applications which can be captured by absolute
sizes and relative ratios of flops, IO bytes and communication bytes.

The second facet captures applications with {\bf important data
  sources with distinctive features}, representative examples of the
data sources include, (i) SQL based, (ii) NOSQL based, (iii) other
enterprise data systems, (iv) set of files (as managed in iRODS), (v)
Internet of Things, (vi) streaming, (vii) HPC simulations, and (viii)
temporal features -- for in addition to the system issues, there is a
temporal element before data gets to compute system, e.g., there is
often an initial data gathering phase which is characterized by a
block size and timing. Block size varies from month (remote sensing,
seismic) to day (genomic) to seconds (real time control, streaming).

The third facet contains {\bf Ogres themselves classifying core
  analytics and kernels/mini-applications/skeletons}, with
representative examples (i) recommender systems (collaborative
filtering) (ii) SVM and linear classifiers (Bayes, Random Forests),
(iii) outlier detection (iORCA) (iv) clustering (many methods), (v)
PageRank, (vi) LDA (Latent Dirichlet Allocation), (vii) PLSI
(probabilistic latent semantic indexing), (viii) SVD (singular value
decomposition), (ix) MDS (multidimensional scaling), (x) graph
algorithms (seen in neural nets, search of RDF Triple stores), (xi)
neural networks (deep learning), (xii) global optimization
(variational bayes), (xiii) agents, as in epidemiology (swarm
approaches) and (xiv) GIS (Geographical Information Systems).

\section{Architecture and Abstractions: HPC and ABDS
  Ecosystems}

\begin{figure*}[t]
	\centering
		\includegraphics[width=\textwidth]{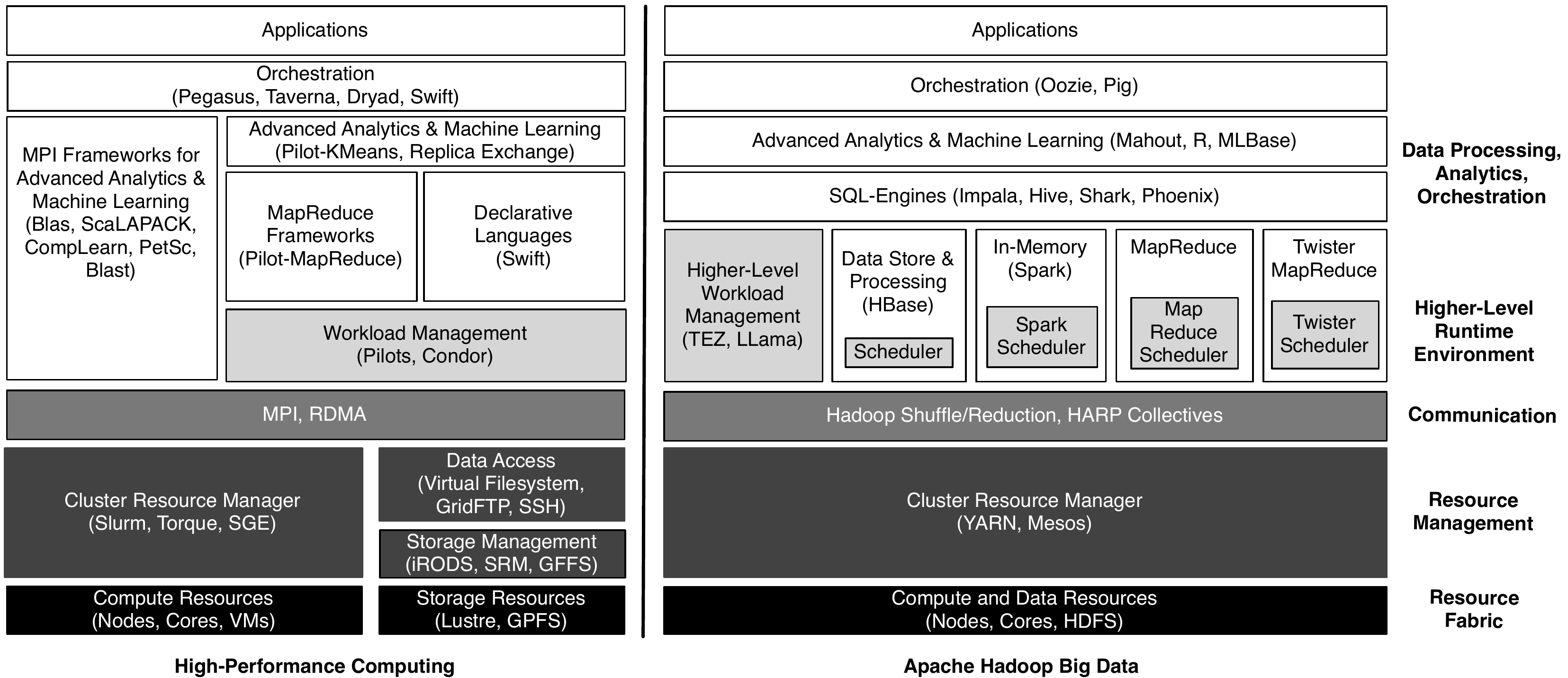}
                \caption{\textbf{HPC and ABDS architecture
                    and abstractions:} The HPC approach 
                  historically separated data and compute;
                  ABDS co-locates compute and data. The YARN
                  resource manager heavily utilizes multi-level,
                  data-aware scheduling and supports a
                  vibrant Hadoop-based ecosystem of data processing,
                  analytics and machine learning frameworks. Each
                  approach has 
                  a rich, but hitherto distinct resource management
                  and communication capabilities.
                  \upp}
	\label{fig:figures_stack}
\end{figure*}

In this section we compare and contrast the ABDS and HPC
ecosystems, viz. the underlying architectural assumptions, the primary
abstractions (both conceptual and implementation).
Figure~\ref{fig:figures_stack} depicts the different layers and
architectural design approaches and highlights some of the primary
abstractions.  For the purpose of our comparison we identified five
layers: resource fabric, resource management, communication,
higher-level runtime environment and data processing/analytics. HPC
infrastructure were traditionally built for scientific applications
aimed toward high-end computing capabilities (small input, large
output). Hadoop in contrast was built to process large volumes of data
(large input, small output), resulting in different software stacks.

\subsection{High Performance Computing}

\note{How does an idealized stack look like? Integration and
  interoperability track?  3 Tracks: Integration, Interop, Application
  Uptake Should MR be in HPC stack?, How should the integrated HPC and
  BigData stack look like?, show the variety of HPC stack!  }

In a typical HPC cluster compute and data infrastructures are
separated: A high-end compute environment -- typically a shared
nothing many-core environment (potentially adding GPUs or other
accelerators such as the Xeon Phi) -- is complemented by a storage
cluster running Lustre 
GPFS~\cite{Schmuck:2002:GSF:1083323.1083349} or another parallel
filesystem connected by a high-bandwidth, low-latency network. While
this meets the need for compute-intensive applications, for
data-intensive applications data needs to be moved
across the network, which represents a potential bottleneck.  Compute
resources are typically managed by a local resource management system
such as SLURM, Torque or SGE. Generally, these systems focus on
managing compute slots (i.\,e. cores).

Storage resources in HPC are shared resources, where a quota is applied on the
data size, but not on I/O. Data locality and other scheduling constraints are
typically not considered. Lustre and GPFS storage resources are typically
exposed as shared filesystem on the compute nodes. In addition several
specialized higher-level storage management services, such as
SRM~\cite{srm-ogf}, \irods~\cite{Rajasekar:2010:IPI:1855046} have emerged.
\irods is a comprehensive distributed data management solution designed to
operate across geographically distributed, federated storage resources. \irods
combines storage services with services for metadata, replica, transfer
management and scheduling. In contrast to ABDS, data-management in HPC is
typically done using files and not using higher level abstractions as found in
the Hadoop ecosystem.

Various other approaches for supporting data-intensive applications on the HPC
infrastructures emerged. For example, different MapReduce implementations for
HPC have been proposed: MPI-based MapReduce implementations, such as
MapReduce-MPI~\cite{Plimpton:2011:MML:2286659.2286704}, can efficiently utilize
HPC features, such as low-latency interconnects and one-sided and non-blocking
communications~\cite{Hoefler:2009:TEM:1612208.1612248}. Further, various
non-MPI MapReduce implementations have been proposed:
Twister/Salsa~\cite{Ekanayake:2010:TRI:1851476.1851593} to support iterative
machine learning workloads,
Pilot-MapReduce~\cite{Mantha:2012:PEF:2287016.2287020} to support
geographically distributed data, etc.

In addition several runtime environments for supporting heterogeneous, loosely 
coupled tasks, e.\,g. \pilotjobs~\cite{pstar12}, many 
tasks~\cite{many-taskcomputing} and 
workflows~\cite{Deelman:2009:WEO:1507767.1507921}.
Pilot-Jobs generalize the concept of a placeholder to provide multi-level
and/or application-level scheduling on top of the system-provided schedulers.
With the increasing importance of data, \pilotjobs are increasingly used to
process and analyze large amounts of
data~\cite{review_sc_pilotdata_2013,pstar12}. In general, one can distinguish
two kinds of data management: (i) the ability to stage-in/stage-out files from
another compute node or a storage backend, such as SRM and (ii) the
provisioning of integrated data/compute management mechanisms. An example for
(i) is Condor-G/Glide-in~\cite{glidein}, which provides a basic mechanism for
file staging and also supports access to SRM storage.
DIRAC~\cite{Tsaregorodtsev:2010zz} is an example of a type (ii) system
providing more integrated capabilities.

\subsection{ABDS Ecosystem}

Hadoop was originally developed in the enterprise space (by Yahoo!) and
introducing an integrated compute and data infrastructure. Hadoop provides an
open source implementation of the MapReduce programming model originally
proposed by Google~\cite{mapreduce}. Hadoop is designed for cheap commodity
hardware (which potentially can fail), co-places compute and data on the same
node and is highly optimized for sequential reads workloads. With the uptake of
Hadoop in the commercial space, scientific applications and infrastructure
providers started to evaluate Hadoop for their purposes. At the same time,
Hadoop evolved with increasing requirements (e.\,g.\ the support for very
heterogeneous workloads) into a general purpose cluster framework borrowing
concepts existing in HPC.

Hadoop-1 had two primary components (i) the Hadoop
Filesystem~\cite{Shvachko:2010:HDF:1913798.1914427} -- an open source
implementation of the Google Filesystem
architecture~\cite{ghemawat2003google} -- and (ii) the MapReduce framework
which was the primary way of parallel processing data stored in HDFS. However,
Hadoop saw a broad uptake and the MapReduce model as sole processing
model proved insufficient. The tight coupling between HDFS, resource
management and the MapReduce programming model was deemed to be too
inflexible for the usage modes that emerged. An example of such a deficit is 
the lack of support for iterative computations (as often found in machine
learning). With the introduction of Hadoop-2 and YARN~\cite{yarn-paper} as
central resource manager, Hadoop clusters can now accommodate any
application or framework.
As shown in Figure~\ref{fig:figures_stack} (right) a vibrant ecosystem of
higher-level runtime environments, data processing and machine learning 
libraries emerged on top of resource fabric and management layers, i.\,e.\ HDFS 
and YARN. Historically, MapReduce was the Hadoop
runtime layer for processing data; but, in response to application
requirements, runtimes for record-based, random-access data
(HBase~\cite{Borthakur:2011:AHG:1989323.1989438}), iterative processing
(Spark~\cite{Zaharia:2012:RDD:2228298.2228301}, TEZ~\cite{tez},
Twister~\cite{Ekanayake:2010:TRI:1851476.1851593}), stream (Spark Streaming)
and graph processing (Apache Giraph~\cite{giraph}) emerged. A key enabler for
these frameworks is the YARN support for multi-level scheduling, which enables
the application to deploy their own application-level scheduling routines on
top of Hadoop-managed storage and compute resources. While YARN manages the
lower resources, the higher-level runtimes typically use an application-level
scheduler to optimize resource usage for the application. In contrast to HPC,
the resource manager, runtime system and application are much more tighter
integrated. Typically, an application uses the abstraction provided by the
runtime system (e.\,g.\ MapReduce) and does not directly interact with resource
management.

Spark is a runtime for iterative processing; it is based on a
Scala-based API for expressing parallel dataflow on top of in-memory,
distributed datasets. For this purpose, Spark introduces resilient
distributed datasets (RDD) as higher-level API that enables
application to load a dataset into the memory of a set of cluster
nodes. The runtime system of Spark automatically partitions the data
and manages data locality during application execution.

Many analytical applications -- in particular in enterprise environments --
rely on SQL as data query language (second Ogre). This led to the development
of several SQL engines (data processing layer): Google's Dremel~\cite{36632},
Hive~\cite{hive}, HAWQ~\cite{hawq}, Impala~\cite{impala} and
Shark~\cite{Xin:2013:SSR:2463676.2465288} are prominent examples. While Hive
was originally implemented based on the MapReduce model, the latest version
relies on TEZ as runtime layer. Similarly, Shark relies on Spark as runtime.
Other SQL engines, such as Impala and HAWQ, provide their own runtime
environment and do not rely on MapReduce or Spark. In addition, hybrid
relational database/Hadoop engines have been proposed:
HadoopDB~\cite{Abouzeid:2009:HAH:1687627.1687731} e.\,g.\ deploys a PostgreSQL
database on every node on which it distributes data using hash partition.
Further, it can access data from HDFS via external tables.
Polybase~\cite{DeWitt:2013:SQP:2463676.2463709} is a similar approach.

Advanced Analytics \& Machine Learning layer applications/frameworks
typically require multiple iterations on the data. While traditional
in-memory, single-node tools, such R~\cite{r-lang} or
Scikit-Learn\cite{scikit-learn} provide powerful implementations of
many machine learning algorithms, they are mostly constrained to a
single machine.  To overcome this limitation, several approaches that
rely on scalable runtime environments in the Hadoop ecosystem have
been proposed, e.\,g.\ Apache Mahout~\cite{apache-mahout} or
RHadoop~\cite{rhadoop}. There are well-known limitations of
Hadoop-1 with respect to support for iterative MapReduce
applications~\cite{Ekanayake:2010:TRI:1851476.1851593}. Thus,
increasingly, iterative runtimes are used for advanced analytics.
MLBase~\cite{DBLP:conf/cidr/KraskaTDGFJ13} is a machine learning
framework based on Spark as lower-level data processing
framework. SparkR~\cite{spark-r} allows R applications to utilize
Spark.

\note{Comparing HPC and HTC components: (i) what is different in the
abstraction level? (ii) comparing implementation (storage, resource management
(scheduling as part of resource management), runtime level,
programming/language level, communication-level).}

\note{planning (part of scheduling, provisioning aspect (VMs startup and 
install) vs. scheduling}

\subsection{Resource Management}
\label{sec:hadoop_rms}

Table~\ref{tab:scheduler} summarizes different architectures for resource 
management: centralized, multi-level and decentralized. HPC schedulers are 
centralized and designed
for rigid applications with constant resource requirements. HPC applications,
such as large, monolithic simulations spawning a large number of cores,
typically utilize tightly-coupled, often MPI-based parallelism. MPI
applications are tightly-coupled and highly latency sensitive. While these
applications have fixed resource demands (with respect to number of cores,
memory and walltime), which does not change during the lifetime, they need to
be scheduled in a way that they simultaneously execute on a system e.\,g.\
using Gang scheduling. Scheduling is done on job-level, i.\,e.\
application-level tasks (e.\,g.\ the execution of the individual processes on
the compute nodes) are not exposed to the resource manager. Thus, scheduling
heterogeneous workload consisting of small, short-running tasks and longer
running batch-oriented tasks represents a challenge for traditional monolithic,
centralized cluster scheduling systems. Often, \pilotjobs are used to overcome
the flexibility limitations and support dynamic applications comprising of
heterogeneous tasks. Data locality is not a primary concern in HPC.

Data-intensive workloads in contrast can be decomposed
into fine-grained, loosely-coupled parallel tasks. By scheduling on task-level
rather than on job-level, the utilization of resources and fairness can be
improved~\cite{Ousterhout:EECS-2013-29}. Multi-level schedulers, such as
YARN~\cite{yarn-paper} and Mesos~\cite{Hindman:2011:MPF:1972457.1972488},
efficiently support data-intensive workloads comprised a data-parallel,
loosely-coupled tasks and allow the dynamic allocation and usage of resource
through application-level scheduling. Decentral schedulers aim to address
scalability bottlenecks and low latency requirements of interactive workloads.
Google's Omega~\cite{omega} and Sparrow~\cite{Ousterhout:EECS-2013-29}, an
application-level scheduler for Spark, are examples of decentral schedulers. In
the following we focus on investigate the evolution of the Hadoop schedulers 
focusing on the central and multi-level approaches.

\begin{table}[t]
    \scriptsize
	\centering
\begin{tabular}{|p{1.3cm}|p{1.9cm}|p{1.95cm}|p{1.9cm}|}
	\hline
    &\textbf{Centralized}     &\textbf{Multi-Level} 		&\textbf{Decentralized} \\\hline
Examples   &Torque, SLURM        &YARN, Mesos, Pilots		  &Omega, Sparrow\\ 
\hline
Workloads &large, parallel jobs &medium-sized tasks &fine-grained tasks \\ 
\hline
Latency &high                    &medium - low &low	\\
\hline
Application Integration  &Submission only   &Application-Level 
Scheduling      &Application-Level Scheduling \\
\hline
\end{tabular}
\caption{{\footnotesize Scheduler Architectures:\label{tab:scheduler}  
centralized, monolithic resource managers are being replaced with more scalable architectures emphasizing application-level scheduling.}\upp\upp\upp\upp\upp\upp}
\end{table}

Hadoop-1 utilizes a centralized scheduling approach using the Job Tracker as
resource manager. Not only represented the Job Tracker a scalability
bottleneck, it also tightly coupled the MapReduce framework significantly
constraining flexibility. In particular in the early days this was not an
issue: Hadoop was often used on top of HPC clusters using Hadoop on
Demand~\cite{hod}, SAGA Hadoop~\cite{saga-hadoop} or MyHadoop~\cite{myhadoop}
or in clouds (Amazon's Elastic MapReduce~\cite{emr} or Microsoft's
HDInsight~\cite{hdinsight}). A limitation of these approaches is the lack of
data locality and thus, the necessity to initially move data to the HDFS
filesystem before running the computation.

Despite the limitations of Hadoop-1, many different resources usage modes for
Hadoop clusters emerged. However, with the increasing size and variety of
frameworks and applications, the requirements with respect to resource
management increased, e.\,g.\ it became a necessity to support batch, streaming
and interactive data processing. Often, higher-level frameworks, such as HBase
or Spark, were deployed next to the core Hadoop daemons making it increasingly
difficult to predict resource usage and thus, performance.
YARN~\cite{yarn-paper}, the core of Hadoop-2, was designed to address this need
and to efficiently support heterogeneous workloads in larger Hadoop
environments. Another multi-level scheduler for Hadoop is 
Mesos~\cite{Hindman:2011:MPF:1972457.1972488}; While there are some mostly 
syntactic differences -- Mesos e.\,g.\ uses a resource offer model, while YARN 
uses resource requests -- it is very similar to YARN providing multi-level 
scheduling for heterogeneous workloads.

An increasingly larger ecosystem evolved on top HDFS and YARN. As
shown in Figure~\ref{fig:figures_stack} applications frameworks
typically rely on a runtime system that embeds an application-level
scheduler that tightly integrates with YARN (e.\,g.  
MapReduce, Spark and HBase all provide their application-level scheduler). 
Increasingly, common requirements are integrated into higher-level, shared
runtime systems frameworks, e.\,g.\ the support for long-running or interactive
applications or multi-stage applications using DAGs (directed acyclic
graph). For example, Llama~\cite{llama} offers a long-running
application master for YARN application designed for the Impala SQL
engine. TEZ~\cite{tez} is a DAG processing engine
primarily designed to support the Hive SQL engine.

As described, typical data-intensive workloads consist of short-running,
data-parallel tasks; By scheduling on task-level instead on job-level,
schedulers, such as YARN, are able to improve the overall cluster utilization
by dynamically shifting resources between application. The scheduler can
e.\,g.\ easily remove resources from an application by simply waiting until
task completion. To enable this form of dynamic resource usage, YARN requires a
tighter integration of the application. The application e.\,g.\ needs to
register an Application Master process, which subscribes to a set of defined
callbacks. The unit of scheduling is referred to as a container. Containers are
requested from the Resource Manager. In contrast to HPC schedulers, the
Resource Manager does not necessarily return the requested number of resources,
i.\,e.\ the application is required to elastically utilize resources as they
become allocated by YARN; Also, YARN can request the de-allocation of
containers requiring the application to keep track of currently available 
resources.

\begin{table*}[t]
    \centering
    \scriptsize
\begin{tabular}{|l|p{1.6cm}|p{2.1cm}|p{2.1cm}|p{1.7cm}|p{1.3cm}|p{3.5cm}|}
\hline
\textbf{Implementation}        &\textbf{Execution Unit}          &\textbf{Data Model }              &\textbf{Intermediate Data Handling} &\textbf{Resource Management} &\textbf{Language}  &\textbf{Hardware}\\ \hline \hline
                
\textbf{(A.1) Hadoop}  &Process &Key, Value pairs (Java Object) &Disc/Local (and network) &YARN &Java &HPC Madrid: 16 cores/node, 16 GB memory, GE \\ \hline

\textbf{(A.2) Mahout}  &Process &Mahout Vectors &Disc (and network) &Hadoop Job Tracker &Java &EC2: cc1.4xlarge, 16 cores, 23\,GB memory, 10GE  \\ \hline \hline

\textbf{(B) MPI}     &Process (long running)  &Primitive Types, Arrays &Message Passing (network) &Amazon/ mpiexec  &C &EC2: cc1.4xlarge, 16 cores/node, 23\,GB memory, 10GE \\ \hline \hline

\textbf{\parbox[t]{2.2cm}{(C.1) Python-Script\\(Pilot-KMeans)}} &Process &Key/Value (Text) &Disk/Lustre  &Pilots, SLURM  &Python, Java  &HPC Stampede: 16 cores/node, 32 GB memory, Infiniband\\ \hline

\textbf{(C.2) HARP}    &Thread (long running)   &Key/Value (Java Object) &Collectives (network) &YARN &Java &HPC Madrid: 16 cores/node, 16 GB memory, GE \\ \hline

\textbf{(C.3) Spark}   &Thread &Key/Value (RDD) &Spark Collectives (network)   &YARN         &Java, Scala &EC2: cc1.4xlarge, 16 cores/node, 23\,GB memory, 10GE \\ \hline

\end{tabular}
\caption{K-Means -- Comparison of different Implementations and Infrastructures  \label{tab:kmeans}\upp\upp\upp\upp\upp\upp}
\end{table*}

\subsection{High-Performance Big Data Stack: A Convergence of
  Paradigms?}
\label{sec:hadoop_hpc}

While HPC and Hadoop were originally designed to support different
kinds of workloads: high-end, parallel computing in the HPC case
versus cheap data storage and retrieval in the Hadoop case, a
convergence at many levels can be observed.  Increasingly, more
compute-demanding workloads are deployed on Hadoop clusters, while
more data-parallel tasks and workflows are executed on HPC
infrastructures. With the introduction of YARN and Mesos, the Hadoop
ecosystem has matured to support a wide range of heterogeneous
workloads.  At the same time a proliferation of tools (e.\,g.\
\pilotjobs) to support loosely-coupled, data-intensive workloads on
HPC infrastructures emerged.  However, these tools often focus on
supporting large number of compute tasks or are constraint to specific
domains; thus, they do not reach the scalability and diversity of the
Hadoop ecosystem.

The ABDS  ecosystem provides a wide-range of
higher-level abstractions for data storage, processing and analytics
(MapReduce, iterative MapReduce, graph analytics, machine learning
etc.) all built on top of extensible kernels: HDFS
and YARN. In contrast to HPC schedulers, a first-order design
objective of YARN is the support for heterogeneous workloads using
multi-level, data-aware scheduling. For this purpose, YARN requires a
higher degree of integration between the application/framework and the
system-level scheduler than typical HPC schedulers. Instead of a
static resource request prior to the run, YARN applications
continuously request and return resources in a very fine-grained way,
i.\,e.\ applications can optimize their resource usage and the overall
cluster utilization is improved. 

While YARN is currently not an option for HPC resource fabrics, trends
and demonstrated advantages have lead to proposals for integrating
Hadoop/YARN and HPC. 
The following
aspects need to be addressed: (i) integration with the local resource
management level system, (ii) integration with HPC storage resources
(i.\,e.\ the shared, parallel filesystem) and (iii) the usage of
high-end network features such as RDMA \pmnote{Is it too popular to define its full form?} and efficient abstractions
(e.\,g.\ collective operations) on top of these.

\emph{Resource Management Integration:} To achieve integration with
the native, system-level resource management system, the Hadoop-level
scheduler can be deployed on top of the system-level
scheduler. Resource managers, such as Condor and SLURM, provide Hadoop
support. Further, various third-party systems, such as
SAGA-Hadoop~\cite{saga-hadoop}, JUMMP~\cite{6702650} or
MyHadoop~\cite{myhadoop}, exist. A main disadvantage with this
approach is the loss of data-locality, which the system-level
scheduler is typically not aware of. Also, if HDFS is used, data first
needs to be copied into HDFS before it can be processed, which
represents a significant overhead. Further, these systems use Hadoop
in a single user mode; thus, cluster resources are not used in an
optimal way.

\emph{Storage Integration:} Hadoop provides a pluggable filesystem
abstraction that interoperates with any POSIX compliant
filesystem. Thus, most parallel filesystems can easily be used in
conjunction with Hadoop; however, in these cases the Hadoop layer will
not be aware of the data locality maintained on the parallel
filesystem level, e.\,g.\ Intel supports Hadoop on top of
Lustre~\cite{lustre_hadoop}, IBM on GPFS~\cite{IBM:2011:UBD:2132803}.
Another optimization concerns the MapReduce shuffling phase that is
carried out via the shared filesystem~\cite{mr-lustre}.  Also, the
scalability of these filesystem is usually constraint compared to HDFS,
where much of the data processing is done local to the compute
avoiding data movements across the network. Thus, HDFS is less reliant
on fast interconnects.

\emph{Network Integration:} Hadoop was designed for Ethernet
environments and mainly utilizes Java sockets for
communications. Thus, high-performance features such as RDMA are not
utilized. To address this issue, RDMA support in conjunction with
several optimizations to HDFS, MapReduce, HBase and other components
for Infiniband or 10 Gigabit networks has been
proposed~\cite{Wang:2011:HAT:2063384.2063461}.
\pmnote{I feel, its a sudden introduction to HARP or Harp does provide some network integration aspects to Hadoop?}\alnote{moved HARP discussion}

\upp
\section{Experiments}
\label{sec:exp}
\upp

\pmnote{Can we add some strong reason/introduction why we choose KMeans?That would address one of the reviewers comment }\alnote{refined}

In the following we run different implementations of the K-Means clustering
algorithm -- an example of Ogre (iii) defined in section II (global machine
learning with parallel computing across multiple nodes). While K-Means the
algorithm is well-understood, a distributed implementation requires efficient
and scalable abstractions for data, compute and network communication. Thus, it 
serves as a good example to analyze these aspects in the different implementations.
Table~\ref{tab:kmeans} summarizes these different implementations. We
categorize these into three categories: (A) Hadoop, (B) HPC and (C) hybrid
implementations. For (A) we investigate (A.1) an Hadoop MapReduce
implementation and (A.2) Apache Mahout~\cite{apache-mahout}; for (B) an
MPI-based K-Means implementation~\cite{liao-kmeans}. We examine the following
hybrid approaches: (C.1) Python Scripting implementation using
\pilots~\cite{Mantha:2012:PEF:2287016.2287020} (\pilot-KMeans), (C.2) a Spark
K-Means~\cite{spark-kmeans} and (C.3) a HARP
implementation~\cite{Zhang:2013:HPC:2523616.2525952}. HARP introduces an 
abstraction for collective operations within Hadoop 
jobs~\cite{Zhang:2013:HPC:2523616.2525952}. While (C.1) provides an
interoperable implementation of the MapReduce programming model for HPC
environments, (C.2) and (C.3) enhance Hadoop for efficient iterative
computations and introduce collective operations to Hadoop environments.

We use Amazon EC2, the Madrid YARN/Hadoop cluster and the Stampede
clusters (which is part of XSEDE~\cite{xsede}) as the different
resources. On EC2 we utilize the cluster compute instance type, which
provides a HPC-style environment. We utilize Elastic
MapReduce~\cite{amazonemr} for managing the Hadoop cluster in scenario
(A.1) and the \texttt{spark-ec2} tool for scenario (C.3). Madrid uses
YARN as resource manager; SLURM is deployed on Stampede.  We run three
different K-Means scenarios: (i) 1,000,000 points and 50,000 clusters,
(ii) 10,000,000 points and 5,000 clusters and (iii) 100,000,000 points
and 500 clusters. Each K-Means iteration comprises of two phases that
naturally map to the MapReduce programming model: in the map phase the
closest centroid for each point is computed; in the reduce phase the
new centroids are computed as the average of all points assigned to
this centroid. While the computational complexity is defined by the
number of points $\times$ number of clusters (and thereby a constant
in the aforementioned scenarios), the amount of data that needs to be
exchanged during the shuffle phase increases gradually from scenario
(i) to (iii), in proportion to the number of points. \pmnote { I think, this is not evident as the
product points $\times$ number of clusters  remains the constant in the three cases..and how does it 
effect the shuffle phase }\alnote{the output of the map phase is the closest centroid for each point, i.e. for n points there are n closest centroids}

\begin{figure*}[t]
    \centering   
    \upp\upp\upp\upp\upp 
\includegraphics[width=.95\textwidth]{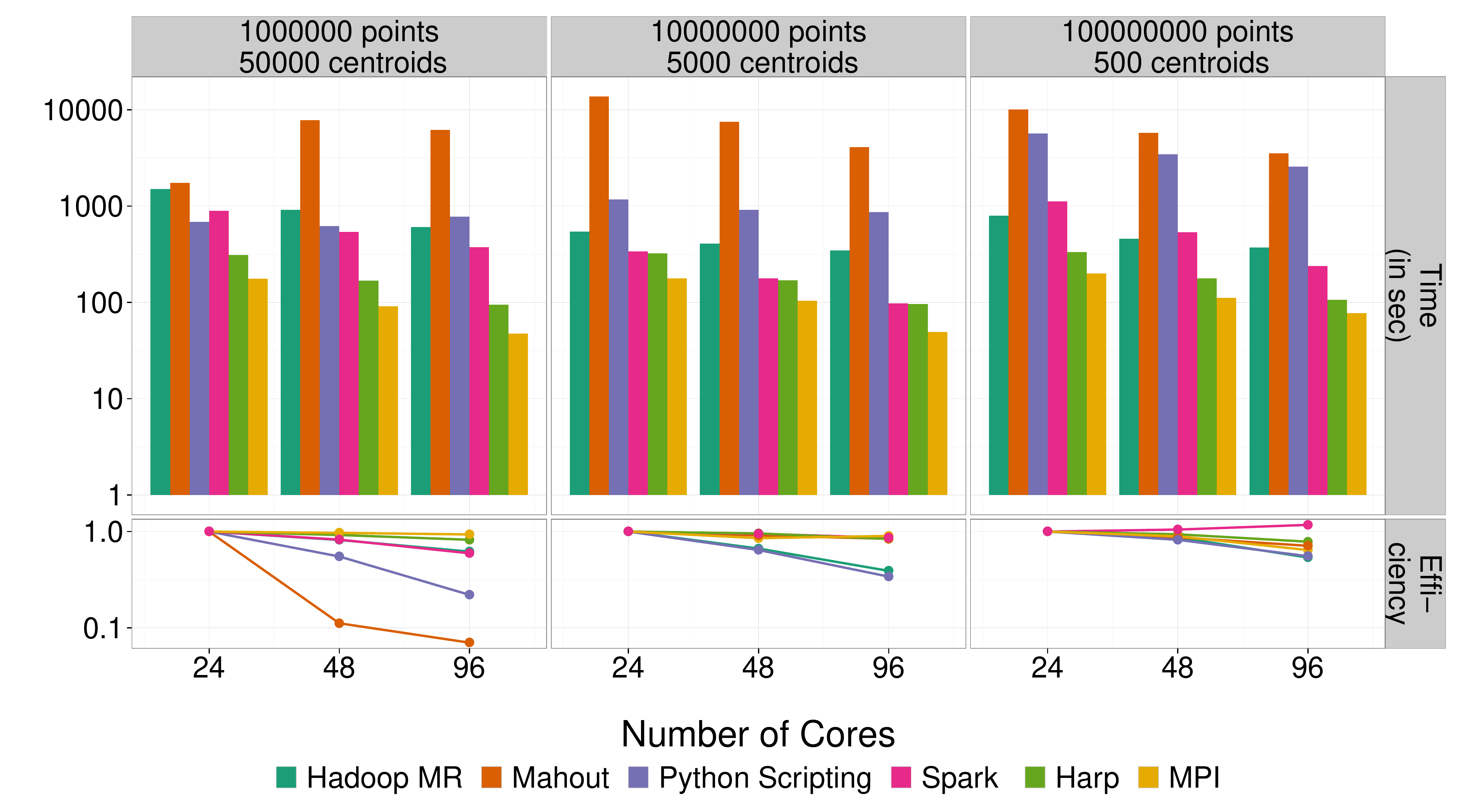}
\upp 
    \caption{Runtime of different K-Means Implementations (y axis in log 
    scale): While MPI 
    clearly outperforms the Hadoop-based implementations, the 
    performance of K-Means can significantly be improved by using hybrid 
    approaches, such as Spark and HARP. By introducing 
    efficient collective and iterative operations known from HPC to Hadoop, the 
    runtime can be improved while maintaining a high-level abstraction for the 
    end-user.\upp\upp\upp\upp\upp\upp}
    \label{fig:experiments_kmeans_kmeans-harp-vs-mahout}
\end{figure*}

Figure~\ref{fig:experiments_kmeans_kmeans-harp-vs-mahout} shows the
results of experiments. Both Hadoop implementations of K-Means (Hadoop
MR (A.1)/Mahout (A.2)) perform significantly worse than the MPI-based
implementation.  The Map Reduce model -- the predominant usage mode of
Hadoop-1 -- has several disadvantages with respect to supporting
iterative machine learning algorithms: The shuffle phase, i.\,e. the
sorting of the output keys and the movement of the data to the reduce
task, is optimized for use cases, such as data grouping, but
introduces a significant overhead where sorting is not needed. In the
case of larger amounts of shuffle data, data needs to be spilled to
disks; sorting is not essential for
K-Means. 
In addition to the inefficiency with each MapReduce, for every
iteration a new job needs to be started, which means that in addition
to the job launching overhead, data needs to be persistent and re-read
to/from HDFS. The MPI implementation in contrast loads all points into
memory once. For communication the efficient collective layer from MPI
(\texttt{MPI\_Allreduce}) is used.

The Python Scripting implementation (C.1) is based on the
Pilot-MapReduce framework, which is an interoperable implementation of
the MapReduce for HPC, cloud and Hadoop environments. The framework
utilizes \pilots for resource managements. For each map and reduce
task, a \cu inside the \pilot is spawned. For data-exchange between
the tasks the shared filesystem (e.g., Lustre on Stampede) is
used. While C.1 outperforms Mahout, it performs significantly worse
than MPI, or other hybrid approaches. In particular for larger amounts
of shuffle traffic (scenario (ii) and (iii)), Hadoop shuffle
implementation is faster. Also, Spark by default compresses the
shuffle data, which improves the performance.

Both Spark and HARP are designed to efficiently support iterative
workloads such as K-Means. While these approaches cannot entirely
reach the performance of MPI, they introduce a unique way of combining
the advantages of MPI with Hadoop/MapReduce. Spark performs slightly
worse than HARP. However, it must be noted that Spark operates at a
higher-level of abstraction, and does not require to operate on
low-level data structures and communication primitives. The RDD
abstraction provides a consistent key/value-based programming model
and provides flexible API for manipulating these. However, since RDD
are designed as immutable entities, data often needs to be copied
in-memory; in each iteration our K-Means implementation generates two
intermediate RDDs. For MPI in contrast only a single copy of the data
is stored in memory and manipulated there.

\upp
\section{Discussion: Convergence of Paradigms}
\upp

Even though a vibrant ecosystem has evolved to support ABDS objective
of providing affordable, scale-out storage and compute on commodity
hardware, the increasing processing/computational requirements, there
is a need for convergence between HPC and ABDS
ecosystem~\cite{bdec14}. Our experiments were designed to expose
important distinctions and relevant considerations for integrated
ecosystem.
 
While our micro-benchmark shows that MPI outperforms the Hadoop-based
implementation, it must be noted that the second generation Hadoop
frameworks, such as Spark, have improved performance significantly by
adopting techniques previously only found in HPC, such as effective
collective operations.  Nonetheless important distinctions remain:
Hadoop-based frameworks still maintain a very high and accessible
level of abstraction, such as data objects, collections etc.,
and are typically written without tight coupling to resource
specifics, e.g., the user can modify some parameters, such as the HDFS
or RDD chunk size, which also controls the parallelism.  In general,
frameworks and tools utilize application-level scheduling to manage
their workloads and provide powerful abstractions for data processing,
analytics and machine learning to the end-user while hiding low-level
issues, such resource management, data organization, parallelism, etc.
HPC applications operate on low-level, communication operations and
application-specific files that often lack a common runtime system for
efficiently processing these data objects.  

Functionalities available in the ABDS ecosystem (more than
110 implementations) typically exceed those available in the HPC
ecosystem, thus reiterating the need for consilience between the
two. Several approaches for convergence
of the two ecosystems have been proposed. Often, these focus on
running Hadoop on top of HPC. However, a lot of the benefits of Hadoop
are lost in these approaches, such as data locality aware scheduling,
higher cluster utilization etc. Thus, we believe that this is not the
right path to interoperability and integration. Furthermore, YARN has
been designed to address the needs of data-intensive applications and
support application-level scheduling for heterogeneous workloads,
there is some ways to go way before YARN can enable both HPC
applications and data-intensive applications on the range of resource
fabrics found in HPC ecosystem.  A possible and promising approach for
interoperability that emerges and will be investigated is the
extension of HPC \pilotjob abstraction to YARN, and the usage of
\pilotdata~\cite{review_sc_pilotdata_2013} for data-locality aware
scheduling.

Our analysis shows that there are technical reasons that drive the
convergence between the HPC and ABDS paradigms, e.\,g.\ rich and
powerful abstractions like collective communications and direct-memory
operations, long the staple of HPC are steadily making their presence
felt in the ABDS. We anticipate the convergence of conceptual
abstractions will soon lead to an integration of tools and technology,
e.g., integration of specific capabilities, especially in the form of
interoperable libraries built upon a common set of abstractions. In
fact, we are working towards such an interoperable library -- Scalable
Parallel Interoperable Data-Analytics Library (SPIDAL) -- that will
provide many of the rich data-analytics capabilities of the ABDS
ecosystem for use by traditional HPC scientific applications. This
will be an incremental but important step towards promoting an
integrated approach -- the high-performance big-data stack (HPBDS) --
that brings the best of both together.

\footnotesize{\noindent\textbf{Author Contributions} -- The
  experiments were designed primarily by AL and JQ, in consultation
  with and input from SJ and GCF. The experiments were performed by
  AL, PM and JQ.  Data was analyzed by all. SJ and GCF determined the
  scope, structure and objective of the paper and wrote the
  introduction, applications and conclusion. AL wrote the bulk of the
  remainder of the paper.

\noindent\textbf{Acknowledgement} This work is primarily funded by NSF 
OCI-1253644.  This work has also been made possible thanks to computer
resources provided by XRAC award TG-MCB090174 and an Amazon Computing
Award to SJ.}

\end{document}